# Whoo.ly: Facilitating Information Seeking For Hyperlocal Communities Using Social Media

**Yuheng Hu**
School of Computer Science
Arizona State University
yuheng@asu.edu

**Shelly D. Farnham**
Microsoft Research
shellyfa@microsoft.com

**Andrés Monroy-Hernández**
Microsoft Research
amh@microsoft.com

**ABSTRACT**
Social media systems promise powerful opportunities for people to connect to timely, relevant information at the hyper local level. Yet, finding the meaningful signal in noisy social media streams can be quite daunting to users. In this paper, we present and evaluate Whoo.ly, a web service that provides neighborhood-specific information based on Twitter posts that were automatically inferred to be hyperlocal. Whoo.ly automatically extracts and summarizes hyperlocal information about events, topics, people, and places from these Twitter posts. We provide an overview of our design goals with Whoo.ly and describe the system including the user interface and our unique event detection and summarization algorithms. We tested the usefulness of the system as a tool for finding neighborhood information through a comprehensive user study. The outcome demonstrated that most participants found Whoo.ly easier to use than Twitter and they would prefer it as a tool for exploring their neighborhoods.

## INTRODUCTION

People rely on multiple sources of information to learn about the communities they live in [28], either for the purpose of community awareness or participation [26]. *Hyperlocal* information is comprised of the news, people, and events that are set within a particular locality, and is of particular interest primarily to the residents of that locality [9]. One of the most important sources of hyperlocal content is social media, such as blogs, microblogs, and social networking sites. Social media has many advantages over traditional media in assisting people's quest for hyperlocal content. With the ubiquity and immediacy of social media, news events often are reported on Twitter or Facebook ahead of traditional news media. For example, the news of both the 2012 Aurora shootings in Colorado and the 2012 Empire State Building shooting in New York City were reported by social media users earlier than by traditional news outlets [8, 2]. Social media has also become one of the few sources of local news—and life-saving information—where traditional media is sometimes censored by governments or even criminal organizations [24]. Moreover, social media has emerged as a dominant platform for communication and connection. As hyperlocal content is mostly generated by and for a community, seamless communication and networking (through one's social networks) can increase exposure to timely peer-generated content, raise people's community awareness, and potentially foster their sense of community [4].

In spite of these benefits, social media tends to be noisy, chaotic, and overwhelming, posing challenges to users in seeking and distilling high quality content from the noise. It should be no surprise that, regardless of the popularity of social media as a source of hyperlocal information, people are still using television and newspapers (among other traditional sources) as their main channels for local information [28]. People need help leveraging social media as a source of information about their hyperlocal communities. At one extreme are the fast-paced, uncurated social media streams: chaotic and overwhelming. At the other extreme are the traditional, authoritative, news sources: slow and less participatory than social media. In this paper, we present Whoo.ly, a novel web service balanced between these two extremes.

Whoo.ly automatically discovers, extracts, and summarizes relevant hyperlocal information contributed on Twitter to facilitate people's neighborhood information-seeking activities. Inspired by the core journalism questions (what, who, where, and when), Whoo.ly provides four types of hyperlocal content in a simple web-based interface (See Figure 1): (i) *active events* (events that are trending in the locality); (ii) *top topics* (most frequently mentioned terms and phrases from recent Twitter posts); (iii) *popular places* (most frequently checked-in/mentioned); and (iv) *active people* (Twitter users mentioned the most).

It is important to note that it is not our goal with Whoo.ly to replace traditional news media. Instead, we want to provide hyperlocal information that is complementary to what both traditional news media and social media have to offer.

The unique features of Whoo.ly are the novel event detection and summarization algorithms we developed. Active neighborhood events are detected using a novel scalable statistical event detector that identifies and groups trending features in Twitter posts. Top neighborhood topics are inferred using a simple yet effective weighting scheme that finds the most important words and phrases from posts. To identify the most popular places in a neighborhood, we used both template-based information extractors and learning-based information extractors. Finally, to distill a ranked list of the active people in a community, we developed a ranking scheme on the social graph of Twitter users based on their mentioning and posting activities.

To evaluate Whoo.ly's utility as a tool for finding neighborhood information, including its user interface and our algorithms, we performed a user study with thirteen residents from three Seattle neighborhoods. Most of our participants believed Whoo.ly provided them with useful neighborhood information, and rated it easier to use than Twitter's native tools.

The contributions of this work are:

- A novel system for discovering hyperlocal information from the social media site Twitter;
- A novel approach for extracting and summarizing trending events from Twitter posts ; and
- Quantitative and qualitative support that our techniques provide higher quality results than existing solutions.

## RELATED WORK

Using new technologies to promote community awareness and participation has long been a research topic for the HCI community and [33, 23]. Web-mediated communities such as Netville and the Blacksburg Electronic Village have demonstrated how the Internet can enhance spatial immediacy, facilitate discussion, and quickly mobilize people around local issues [11, 5].

The prevalence of "Web 2.0" has provided new opportunities for technologies to facilitate better information seeking and communication about local communities. In particular, social media tools have been used to report various activities including breaking news [22], public debates [17], crises like floods [31], earthquakes [29], or even during wartime [24]. Recently, leveraging social media resources for local communities has drawn considerable attention in both research and industry. Such efforts include Livehoods [7] and i-Neighbors [10]. Among them, CiVicinity [14] provides a hyperlocal community portal that integrates information from Facebook, blogs, calendars, and other sources to promote civic awareness and participation. Virtual Town Square (VTS) [20] also aggregates local information from a predefined set of information sources (government, schools, and news organizations) to improve community engagement. Our work uniquely builds on this line of research by exploring automatic solutions to the detection, extraction and summarization of neighborhood information from noisy Twitter posts.

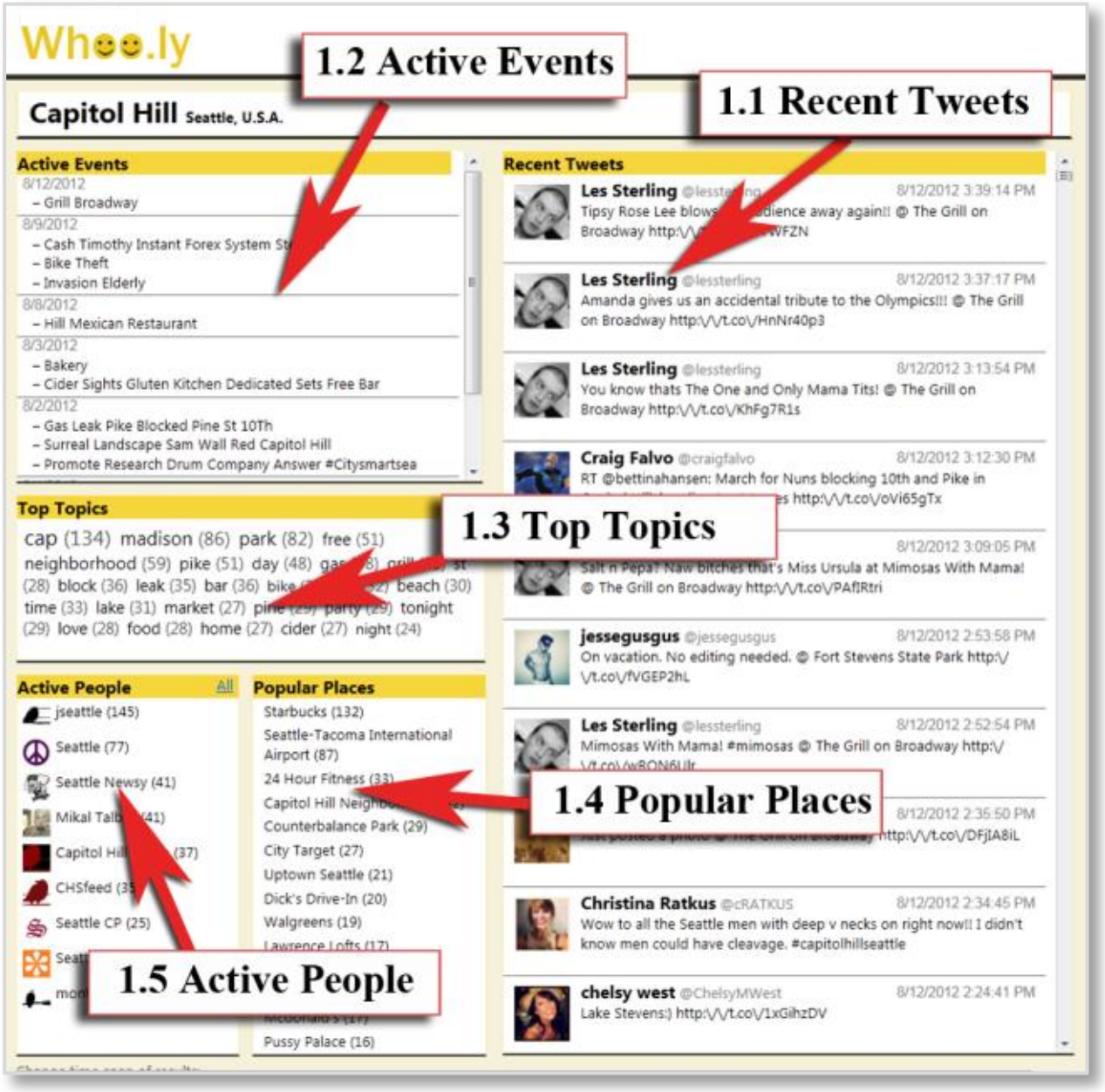

**Figure 1. The main Whoo.ly interface, with the recent Twitter posts and summaries of events, topics, places, and people.**

The hyperlocal content in Whoo.ly is automatically mined from Twitter, which presents unique challenges not directly addressed by related work: (1) Prior solutions on event detection from social media commonly employ the strategy of clustering similar Twitter posts, using a classifier to predict the event-related clusters, and then extracting events from these clusters [1]. Such an approach may work well on long text documents (e.g., blogs) but perform poorly on Twitter posts, since clustering outcomes can be noisy; at the same time, analyzing sparse and short text can be challenging [15]. In addition, this strategy needs to be trained in advance. In contrast, our proposed event detector finds trending events without any supervision and, more importantly, it is highly scalable, making it feasible to efficiently handle large-scale social media data. (2) Our method of finding top topics was inspired by the TF-IDF statistics that assign scores for terms based on their mentioned frequency within and across documents. Even though there are other efforts to find top topics from Twitter posts [27], such approaches often take a long time to run to discover meaningful topics, and we seek to provide reasonably real-time results. (3) Information extraction has been a long-standing research topic [6]. In Whoo.ly, we use a hybrid approach of both template-based and learning-based extractors to find popular places in Twitter posts.

## WHOO.LY OVERVIEW AND DESIGN PROCESS

In this section, we first provide an overview of Whoo.ly and its features. Then, we highlight the motivations underlying the choices we made in the design process.

Whoo.ly is a web service built on top of Twitter. Its goal is to provide people with relevant and reliable hyperlocal news content. By browsing the website, people immediately find what is happening in a specific neighborhood. Whoo.ly

provides four hyperlocal content types: active events, top topics, active people, and popular places (See Figure 1). All of them are automatically extracted and summarized from Twitter using various approaches we developed, such as statistical event detector, graph-based ranking algorithm, and information extractors (see the System Design section for more detail).

Early in the design process for Whoo.ly, we examined local newspapers, community blogs, existing hyperlocal sites, and Twitter. The exploratory study revealed several interesting results that we used to motivate the design of Whoo.ly: (1) The majority of the people only consume information—they do not produce it but only read it; (2) People become more active in reporting and disseminating local breaking events (e.g., shooting, water leak) on Twitter by reposting related tweets; (3) People tend to follow neighborhood curators or bloggers who are dedicated to posting hyperlocal content; and (4) Local media and local news services effectively cover important local topics. However, people further seek hyperlocal content generated by people in their communities.

We performed an additional preliminary analysis of Twitter data to help inform our design decisions, answering the following questions: (1) can we find a base of local Twitter posts based on neighborhoods; (2) were there enough messages to seed a neighborhood website; and (3) what do people care to talk about on Twitter regarding their neighborhoods? We first queried for all Twitter messages from people who claimed Seattle as their home town for the month of October of 2011. We then performed a simple extraction of Twitter messages that mentioned one of 83 Seattle neighborhoods. We found 50,609 unique Seattle users and produced 1.2 million messages (about 8% of total Seattle population), out of which 5% explicitly mentioned Seattle, and another 2% mentioned a Seattle neighborhood. On average 132 people posted per neighborhood over the month, averaging 1.8 messages each, which translates into about 8 messages per day per neighborhood. There was great variability across neighborhoods, but we considered the above averages to be a promising start and used them as the volume of neighborhood Twitter messages to expect.

To examine message content, we sampled 24% of the messages (424) from three neighborhoods pre-selected for being diverse from each other. We first coded the messages for whether they were erroneously assigned to the neighborhood. Surprising, only 21 messages (5%) were erroneously assigned, largely because of overlapping neighborhood names and other place names (e.g., the area “Mount Baker” and the mountain “Mount Baker” it was named after). We then looked at how many were personal in nature, of little interest to anyone aside from the author’s friends. We found that 13% of messages were of this nature. Places check-ins comprised another 55 messages (10%), which we expect might be interesting when aggregated but not at the individual level. Six items were impossible to interpret and were left unclassified. The remaining messages were 71% on topic, meaningfully pertaining to the neighborhood. We further inspected and coded by message type and whether or not they were about a current event. We defined a *current event* as a real-world occurrence with an associated time period such that if it is not observed, experienced, or attended in that time period a person will not be able to do so later. Thus a crime, a fire, a festival, or a Friday happy hour are current events. In contrast, a photo shared online, a news story link, a recommendation to try a restaurant, or a shoutout of thanks are not. We found that 55 % of the remaining Twitter messages were about an event.

| Types of neighborhood messages | Percent |
|---|---|
| Neighborhood Affirmations | 13% |
| Local Business Updates | 11% |
| Local News | 11% |
| Recommendations | 11% |
| Civic Activity | 10% |
| Classified Ads | 9% |
| Social Events | 8% |
| Crime, Fire, Emergency, Road Reports | 7% |
| Deals/Coupon | 7% |
| Talks or Classes | 4% |
| Festival or Outdoor Market | 4% |
| Local Sports | 2% |
| Salutations, Thanks, Shoutouts | 2% |
| Acts of Nature | 1% |

**Table 1: Types of neighborhood messages shared on Twitter.**

All message types in our data sample were classified as depicted by Table 1. Topics such as crime reports, Yelp-like recommendations, and local news were not surprising. The neighborhood affirmations and salutations were surprising, where people in the community post messages talking about how much they love their neighborhood, or community-affirming, humorous messages reinforcing the neighborhood’s stereotypical traits.

Based on these findings, we decided to focus first on detecting events and then to promote community-enabling features such as a list of top users so that people can know and follow each other. To prevent information overload, we also provided top topics so that people can quickly learn the common neighborhood topics in the Twitter posts (tweets).

## DATA COLLECTION OF WHOO.LY

Whoo.ly is built on Twitter. We utilized the Twitter Firehose that is made available to us via our company’s contract with Twitter. Since we are interested in discovering hyperlocal content for local communities in various geographic regions, we needed to obtain a set of Twitter posts from each region. Twitter offers two possible ways to infer a tweet’s location: GPS coordinates associated with a tweet or the user’s location in their profile. In this work we used the location information derived from the user profile since the number of Twitter posts found by GPS coordinates is very limited (about 0.6%). From our preliminary analysis using this method, we found a reasonable quantity of on-topic neighborhood messages.

We observed that most Twitter users prefer to mention only their city instead of local community for the profile location, probably due to privacy concerns [13]. As a result, we first obtained Twitter posts from the Firehose, where each associated user profile location matches one of the dictionary strings for a city, e.g., “Seattle” or “Sea”. Next, we mapped these Twitter posts into different neighborhood regions by matching their textual content against a list of neighborhoods. Note that the neighborhood list for each city is created by domain experts who have comprehensive experience with the neighborhood development and boundaries in that city.

We used a dataset that included about 2.2 million Twitter posts in English from about 120,000 unique users whose profile location indicated they are from Seattle, over a three-month period from June 1, 2012 to Aug 15, 2012. While we mainly used this static dataset for developing our prototype, our methods may easily be extended to handle real-time tweet streams.

## SYSTEM DESIGN OF WHOO.LY

In this section we describe the system design of Whoo.ly (Figure 1), including the interface design of its components and the technical design behind them. Whoo.ly’s interface is implemented in HTML, CSS, and Ajax controls toolkits, served by ASP.net on the cloud service Windows Azure.

Whoo.ly first shows a start page, where a user selects his or her country, city, and neighborhood through drop-down lists. After selecting their location, users are taken to the results page (Figure 1), which displays recent Twitter posts, top topics, popular places, and active people.

### Recent Twitter Posts

Whoo.ly presents recent Twitter posts in a scrolling list on the right side of the results page (Figure 1.1). Each row in this list contains a detailed Twitter profile for a user on the top, and his or her recent posts at the bottom. The profile includes standard elements retrieved from Twitter such as the user name, screen name, user’s profile image, user’s profile location, and the posting time of the messages.

Whoo.ly only provides the most recent Twitter posts from a time window of 14 days mainly because people are usually only interested in most recent Twitter posts. Nevertheless, the length of the time span can be easily adjusted through a drop-down list at the bottom of the results page.

### Active Events

Whoo.ly presents an active events list calendar (Figure 1.2) on the upper left side of the results page. Each entry shows the events organized by date. Every event is summarized by a list of terms and, by clicking on its name, the user is taken to a page (Figure 2) containing all the posts that are about that event, ranked by their relevance score using vector similarity [22].

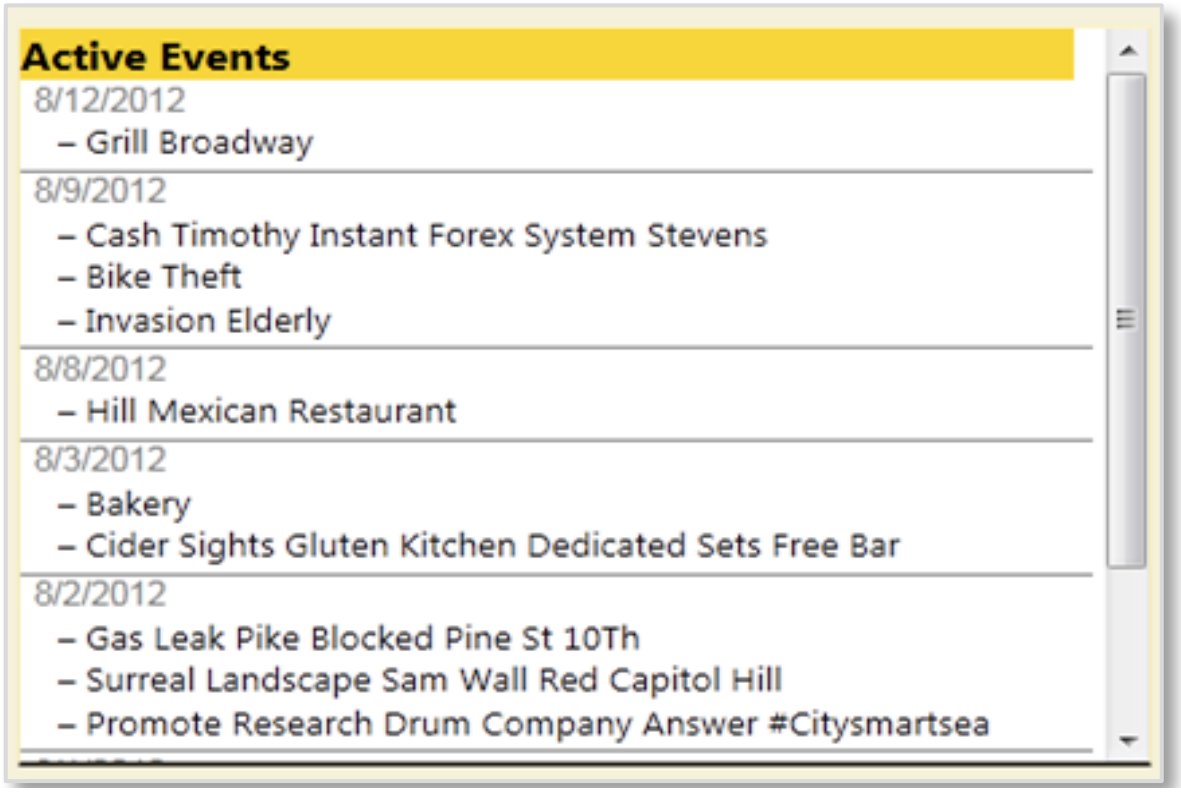

**Figure 2: A close-up view of the Active Events pane. Events are organized by date and are represented by a list of terms most associated with each event.**

A core research question in this component is how can we detect *active events*? Given a tweet stream, where each tweet consists of a set of features $F_1$ , $F_2$ ,... (e.g., gas, leak, danger, etc.), *active event detection* is finding a set of active events, where an active event consists of a set of *topically-related trending* features, at a given time period. To address this question, we developed a novel event detector, which (1) identifies trending features from Twitter posts using trends indicators; and (2) clusters the topically-related trending features into event-clusters using topic modeling and a clustering scheme. Next, we explain our approach in detail.

#### *Trending Features Identification*

To identify trending features from a substantial volume of Twitter posts, we first need to determine what is *trending*. Inspired by the model of theoretical “bursts” in streams of topics [12], we define trending as a time interval over which the rate of change of momentum (i.e., product of mass and velocity) is positive. We further define that *mass* is the current importance of the feature and the *velocity* is the rate of change of the feature’s frequency in Twitter posts, during a time period. Since it is hard to directly measure the momentum from these values, we chose to use the trend analysis tools *EMA* (Exponential Moving Average), *MACD* (Moving Average Convergence Divergence), and *MACD* histogram from the quantitative finance literature [25] to yield established measures of momentum. Next, we explain how these tools work to identify trending features from Twitter posts.

Given a feature $F$ and its time series $S(F) = \{f_1, f_2, \dots f_m\}$, $f_i$ denotes the frequency that $F$ is mentioned by the Twitter posts posted within the $i$-th period. For example, the word “morning” can have a time series $S$ = {248,305,154,52,24,9} from 8 a.m. to 2 p.m. of the day, in which it was mentioned 248 times by the Twitter posts from 8 a.m. to 9 a.m., 305 times from 9 a.m. to 10 a.m., and so on. Moving averages are commonly used with time series data to smooth out short-term fluctuations and highlight longer-term trend. Here, we compute the $n$-hour *EMA* for $S(F)$ as: $EMA_i = \alpha \times f_i + (1-\alpha) \times EMA_{i-1}$, where $\alpha = 2/(n+1)$ is a smoothing factor, $n$ is a time lag, and $1 \le i \le m$ is the index of time period. Essentially, the *EMA* smoothens out noises of $F$ by averaging its time series over a specific number of periods.

Next, to spot changes in the momentum of $F$, we compute the *MACD* statistics, which is defined as the difference between the $n_1$- and $n_2$- hour *EMA* for $S(F)$, where $n_1$ and $n_2$ are time lags. Finally, to identify whether and when $F$ is trending, we need to quantify the rate of change of its momentum. Therefore, we calculate the *MACD* histogram, defined as the difference between $F$'s *MACD* and its signal line (the n-day *EMA* of *MACD*). As this difference measures the rate of change, the result at a given time period can be either positive (indicating $F$ is trending up) or negative (indicating $F$ is trending down).

In some cases, the trending features may occur repeatedly. For example, "morning" can be trending from 8 a.m. to 11 a.m. every day. Such trending feature may be less interesting compared to the ones which are single occurrences. To resolve this problem, we assign a "novelty" score to the identified trending feature according to their deviation from their expected trend. More specifically, for a trending feature $F$, we denote $R(h, d, w, F)$ as its *MACD* histogram result during hour $h$, day $d$, and week $w$. With this notation, we can compare $F$'s trend in a specific day or hour in a given week to the same day or hour in other weeks (e.g., 9 a.m. on Monday, Aug 6, 2012, vs. the trend on other Mondays at 9 a.m.). Let $Mean(h, d, F)$ and $SD(h, d, F)$ denote the *average trend* and the *standard deviation* of $F$ on hour $h$ and day $d$ over week $w_1$ to $w_n$, respectively. Then, the novelty score of feature $F$ on hour $h$, day $d$, and week $w$ is defined as: $Score(h, d, F) = [R(h, d, w, F) - Mean(h, d, F)]/SD(h, d, F)$. Based on this score, we rank each feature to find the novel trending features.

In practice, to detect the daily active events, we first built a dictionary of features from all the Twitter posts of one day. Then, we created a time series for each feature by counting their frequencies in Twitter posts in every two hours. As a result, we have a 12-hour-long time series for every feature. Then, we applied the *EMA*, *MACD*, *and MACD* histogram over the time series data to identify whether and when a feature is trending. Finally, for every two hours, we picked the trending feature which (1) is least mentioned 20 times in the Twitter posts from that time period, and (2) has a novelty score among the top 25 scores for all trending features from that time period. Since these steps are computable in an online fashion [12], our approach is highly efficient.

#### *Trending Feature Clustering*

To group the trending features into topically-related event-clusters, we use the shared nearest neighborhood (SNN) clustering algorithm [18]. We chose this algorithm because it is scalable and does not require a priori knowledge of the number of clusters (as Twitter posts are constantly evolving and new events get added to the stream over time).

The SNN algorithm is executed as follows: each trending feature is a node of the graph and each node is linked to another by an edge if it belongs to the $k$ neighbor list of the second object. Here, we define feature $F_1$ is the *neighbor* of feature $F_2$ only if $F_1$ and $F_2$ are topically-related (e.g., "gas" can be a neighbor to "leak" but may not be to "party"). To learn a feature's topic, we use topic modeling [3], a popular machine learning tool for getting topic distributions from text. In order to measure the topical relationship between two features, we use the Jensen-Shannon divergence on their topic distributions. As a result, if the distance is above a threshold, the two features are neighbors.

### Top Topics

Below the trending events section, Whoo.ly shows a list of top topics (with their frequencies) that are being discussed in the recent Twitter posts (Figure 1.3). Clicking a topic leads to a page showing all the Twitter posts about it. This component helps people quickly understand and familiarize themselves with the most important topics about the neighborhood appearing in Twitter posts. We design a fast approach by applying normalized TF-IDF statistics for each uni-, bi-, and tri-gram from the recent Twitter posts. We then rank these grams to render this component.

### Popular Places

Beyond the event and topics, Whoo.ly shows a list of 15 most popular places (Figure 1.4) that people keep checking into and mentioning in Twitter posts. Similar to other components, clicking a place leads to a page showing all Twitter posts about this place. This component helps people discover interesting places in their neighborhood and learn what is happening there. Extracting these places from Twitter posts requires automated information extraction, which has been a long-standing research topic in NLP and machine learning [6]. In the next sections, we describe two types of extractors we use to build this component, namely a template-based extractor and learning-based extractor.

#### *Template-based Information Extractor*

Through our manual inspection of the Twitter posts content (see the Overview section), we found there is a small percentage of Twitter posts (approximately 7%) that were posted by Foursquare check-ins. Such Twitter posts have a specific template in their content: begin with the phase "I'm at", followed by a place name (e.g., *Space Needle*), and followed by its address (e.g., *400 Broad Street, Seattle, WA 98102*). Given this structure, we designed a template-based extractor using regular expressions to distill the place information.

#### *Learning-based Information Extractor*

For Twitter posts without explicit format for location inference, we used a statistical information extractor. It is built on top of an n-gram language Markov model and previously trained on Wikipedia pages, Tweets, and Yelp data [32]. We apply it to analyze the Twitter posts to extract entities for places, e.g., restaurants, parks, streets, stadiums, etc.

### Active People

Last, Whoo.ly displays a list of top 10 most active people (i.e., Twitter users) for the corresponding neighborhood (Figure 1.5). Each record in the list combines a user's profile and the frequency this user posts or was recently mentioned by other people. In addition, Whoo.ly also presents the profiles, latest Twitter posts, and activities of all the users who have recently posted Twitter messages (by clicking "All" on the up right corner of this division). With this component, one can easily identify who are the active and influential people in the neighborhood and can decide to follow their activity.

To build this component, we developed a PageRank-like algorithm to rank the Twitter users based on their mentioning and posting activities. Specifically, a directed graph $D(V, E)$ is formed with the users and the "follower-followee" relationships among them. $V$ is the vertex set, containing all the users. $E$ is the edge set. There is an edge between two users if there is "following" relationship between them, and the edge is directed from follower to followee. Our algorithm performs an activity-specific *random walk* on graph $D$ to calculate the rank. It visits each user with certain transition probability by following the appropriate edge in $D$. The probability is proportional to a linear combination of the interactions between two users (e.g., RT, mentioning, reply) and how many Twitter posts a user has posted recently. The idea is that the more activities a user has, the higher this user's rank is.

## USER STUDY

We evaluated Whoo.ly as a tool for users to learn about what is happening in their neighborhood using a within-subjects comparison of Whoo.ly and Twitter, where users completed a series of information-seeking tasks for each platform and then provided feedback. For our user study, we focused only on three Seattle neighborhoods for which we were able to recruit participants.

### Participants

We introduced 13 Seattle residents into a private, pre-release version of Whoo.ly through five focus group sessions, with two or three people per session. Participants were recruited from a pre-existing database of people who for the most part had expressed interest in user studies. Potential participants in the database were first filtered for address zip codes in our target neighborhoods. After receiving phone calls to screen for whether they continued to live in the neighborhood and had a Twitter account, they were scheduled to participate in one of five sessions. In exchange for their participation they received their choice of a software gratuity or gift card. Participants were on average 30 years of age (ranging from 23 to 48). 54% of them were female and 46% were male. Ten participants were white, one Asian, one Native American, and one had other ethnic identity. The majority of participants were from the Capital Hill neighborhood (69%), with 23% from Wallingford and 8% from Rainier Valley. These neighborhoods differed in density, SES, and level of existing community infrastructure.

### Procedure

During two-hour user sessions participants first completed a preliminary questionnaire. They then briefly discussed their current communication practices for finding and sharing neighborhood information in a semi-structured focused group. Participants then individually completed a series of tasks with both Whoo.ly and Twitter using laptops with an Internet connection we provided. After a brief discussion of participants' experiences, we ended the session by having them rate a series of Twitter messages for neighborhood content.

#### *Preliminary Questionnaire*

Participants first completed a brief preliminary questionnaire to assess demographic information, use of Internet, and social media, and measure of their current neighborhood including psychological sense of community, neighborhood communication efficacy, and civic engagement. We measured psychological sense of local community [30], or the feeling of connection, belonging, and loyalty to a local community, with items such as *"I feel loyal to the people in my neighborhood*," *"I really care about the fate of neighborhood*," and *"I feel like I belong in my neighborhood*." Civic engagement was measured using items from the Civic Engagement Questionnaire [21], a standard measure asking how often respondents had engaged in various civic activities such as *"Spending time participating in any neighborhood community service or volunteer activity*" and *"playing a leadership role in my neighborhood (such as local government or leadership in a club)."* Neighborhood communication self-efficacy, including communication self-efficacy**,** was measured with items adapted from the California Civic Index [19] that addressed communication, including *"I know how to collect information and be informed about neighborhood issues*," and *"I know how to get in touch with members of my neighborhood when I need to communicate with them."* For each measure, items were rated on a Likert scale of 1 to 7, where 1 = *not at all* and 7 = *extremely so*, and then items were averaged for analysis.

#### *Focus Group*

To further elucidate existing information-seeking and communication practices, we then had participants discuss their neighborhoods using a semi-structured group interview. Participants first described the character of their neighborhoods, how long they have been living there, and whether they had a sense of connection or community to their neighborhood. We then asked participants to discuss what kinds of information they cared to learn about in their neighborhoods. Participants then described the tools they currently use to seek out information or communicate with others around neighborhood issues and where they would like to see changes or improvements in the tools available.

#### *Neighborhood Information Seeking Task*

Following the focus group, participants individually completed a series of four information-seeking tasks, once in Whoo.ly, and once in Twitter. Each participant completed the tasks separately on a laptop with an Internet connection following instructions in a paper packet. The order of the tasks (Whoo.ly vs. Twitter) was counterbalanced across sessions, ending with seven participants completing the Twitter tasks first and six participants completing the Whoo.ly tasks first. Participants were instructed, *"for this part of the study we will have you explore what's happening in your neighborhood using [Twitter or Whoo.ly]."* The four tasks were: 1) find neighborhood events: *"try to find three interesting or significant events that happened in your neighborhood the past couple of weeks"*; 2) find neighborhood reporters: *"imagine you wanted to try to follow three people to help you keep up to date with what's happening in your neighborhood—try to find those three people you would follow"*; 3) find neighborhood topics: *"imagine you wanted to find out what kinds of topics your neighborhood tends to care about—try to find three of these topics;"* and 4) find neighborhood friends: *"imagine you wanted to get to know some people in your neighborhood better—find three people you might want to know more".*

Participants were instructed to spend only a few minutes on each task, to get a sense for the experience in the system they were evaluating. After completing each task participants rated the ease of the task to complete, how confident they felt about their answers, and how engaged they were by the task (that is, to what extent they found it fun or interesting).

Following the completion of these tasks, participants rated the overall usefulness and ease of each system (Twitter and Whoo.ly), the extent to which it provided a good overview of what is happening in their neighborhood, the extent to which it provided a sense of connection, and which system they would prefer to use for finding out what is happening in their neighborhoods. Finally, participants were asked to rank their preference for individual aspects of the Whoo.ly interface and provide opened-ended feedback to questions about what they liked, disliked, and possible improvements.

### *Tweet Rating Task*

In order evaluate the event-detection algorithms, participants were asked to rate a randomly selected series of Twitter posts from the period spanning two weeks prior to that of the current Whoo.ly system. For each tweet, participants rated if it was a about a neighborhood event and if so, how significant was the event to their neighborhood, where 1 = not at all, few people involved, and 7 = extremely so, entire neighborhood involved.

## RESULTS

In analyzing our results, we first examined our participants' existing neighborhood information-seeking and communication practices to better shed light on their experience of Whoo.ly and potential considerations for a real world deployment of this system. We then assessed how well participants completed information-seeking tasks in Whoo.ly, providing a comparison to Twitter as a baseline tool for searching and browsing Twitter messages. Finally, we further examined themes that emerged from participant ratings and discussions that would meaningfully influence the design of Whoo.ly and similar systems.

### Existing Practices

In our preliminary questionnaire participants rated themselves as having high levels of overall Internet experience, with 39% categorizing themselves as intermediate, 45% as advanced, and 16% as expert. Seventy-six percent of participants reported spending four or more hours a day using the Internet. For communicating and sharing with others, participants reported text messaging ($M$ = 6.5, $SD$ = 0.66) and e-mail ($M$ = 6.6, $SD$ = 0.65) to be extremely important, then social networking sites such as Facebook ($M$ = 5.9, $SD$ = 1.00), blogs ($M$ = 4.1, $SD$ = 1.32), Twitter ($M$ = 3.6, $SD$ = 1.90), and mailing lists less so (where 1 = not at all, and 7 = extremely so).

Most of the participants in our study cared very much about their neighborhoods, reporting fairly high levels of psychological sense of community ($M$ = 5.0, $SD$ = 0.83). The few exceptions made apparent from our interviews were individuals new to the neighborhood, or one participant who felt his neighborhood was too transitional by nature to become attached to it. However, the participants had lower levels of civic engagement (3.0, 1.27) and communication self-efficacy ($M$ = 3.8, $SD$ = 1.8).

When asked to what extent they could collect information and be informed about neighborhood issues, participants' responses were on average moderate ($M$ = 3.9, $SD$ = 1.8). An examination of the distribution of this variable suggests it is bimodal, for example people either were low (45% at 2 or 3) or high (39% at 5 or higher) in their ability to find information or communicate with their neighborhood. When participants were asked how exactly they learned about what was happening in their neighborhoods, resources were quite diverse, including local newspapers, local blogs, following business on Twitter, local meetings, Facebook groups, coffee shops, and services such as Reddit, Google, and Yelp. However, local blogs clearly played a prominent role and word of mouth was frequently mentioned as a source of information. Several people mentioned Facebook or Facebook groups, but these were groups of people they knew who were in their neighborhoods, rather than public Facebook groups for the entire neighborhood. Further, it was clear that some neighborhoods had many more resources available than others.

We further asked what kinds of neighborhood information participants wanted to know about. Emerging themes were events such as local festivals and block parties, crime, new restaurants and bars, building developments, people, and local business promotions such as happy hours and coupons. Events and crime were most frequently mentioned, particularly as they impacted the local community. One participant's response was,

> *Community stuff—like I heard about neighborhood night out but I didn't know about it, my street closed and people were out drinking and barbecuing and I didn't know about it—you know about the big things, but little community stuff, that stuff you should know.*

On average, participants were not confident they knew how to get in touch with members of their neighborhood when they needed to communicate with them ($M$ = 3.5, $SD$ = 1.9). When participants were asked, if they needed to communicate with members of their neighborhood community about neighborhood issues, how would they do so, face-to-face was rated the most highly ($M$ = 5.2, $SD$ = 1.8), followed by Facebook groups. During the interviews across sessions participants similarly exhibited low confidence in how they would go about communicating with their neighbors, and expected they would resort to walking down the street. One participant replied, *"old fashioned way, knock on door. Too many people in the neighborhood to have phone numbers and e-mails."* More tech-savvy participants said they would contact the local blog or access their neighbors' e-mail addresses.

We asked participants to discuss their Twitter usage in particular, given the focus of Twitter as a source of public information in Whoo.ly. All participants had an account, but the majority used it primarily to consume information, either the news or their friends' posts. Only a few used Twitter to follow their neighborhood bloggers or neighborhood businesses.

To summarize, we found that our participants were fairly tech-savvy and felt fairly attached to their neighborhoods. While only a few were more civically engaged, most reported they would want to be more so. However, the participants did not have a strong sense for how to find out about what was happening in their neighborhoods or how to get involved. Particularly, they were not sure how they would go about communicating with others in their neighborhood about issues they cared about. Participants were especially interested in learning about local community events and crimes and relied heavily on one or two hyperlocal bloggers to do so.

### Whoo.ly Evaluation

Participants completed four tasks exploring their neighborhood—find recent events, find local neighborhood reporters, find neighborhood topics, and find potential neighborhood friends—using both Whoo.ly and Twitter. We performed an omnibus repeated measures ANOVA (technology X task X type of rating) to test for the impact of type technology across measures of task ease, confidence in completing task, and task engagement. Overall, we found a significant effect of technology ($F$(1,11) = 3.02, $p$ = 0.05 1-tailed[1]), with participants showing preference for Whoo.ly. As can be seen from Figure 3, people overall found Whoo.ly easy to use and found the tasks easier to complete in Whoo.ly than in Twitter. We found neighborhood communication self-efficacy to be a meaningful covariate interacting with this effect ($F$(1,11) = 3.3, $p$ = 0.04, for interaction of technology X task X self-efficacy), meaning participants with lower levels of self-efficacy were likely to favor Whoo.ly over Twitter, especially for the find-friends task.

These results suggest that Whoo.ly is particularly easy for users to learn more about their neighborhood if they do not already have effective tools to find information and access people in their neighborhood.

[1] Given the small N and a priori predictions, we report 1-tailed $p$ values.

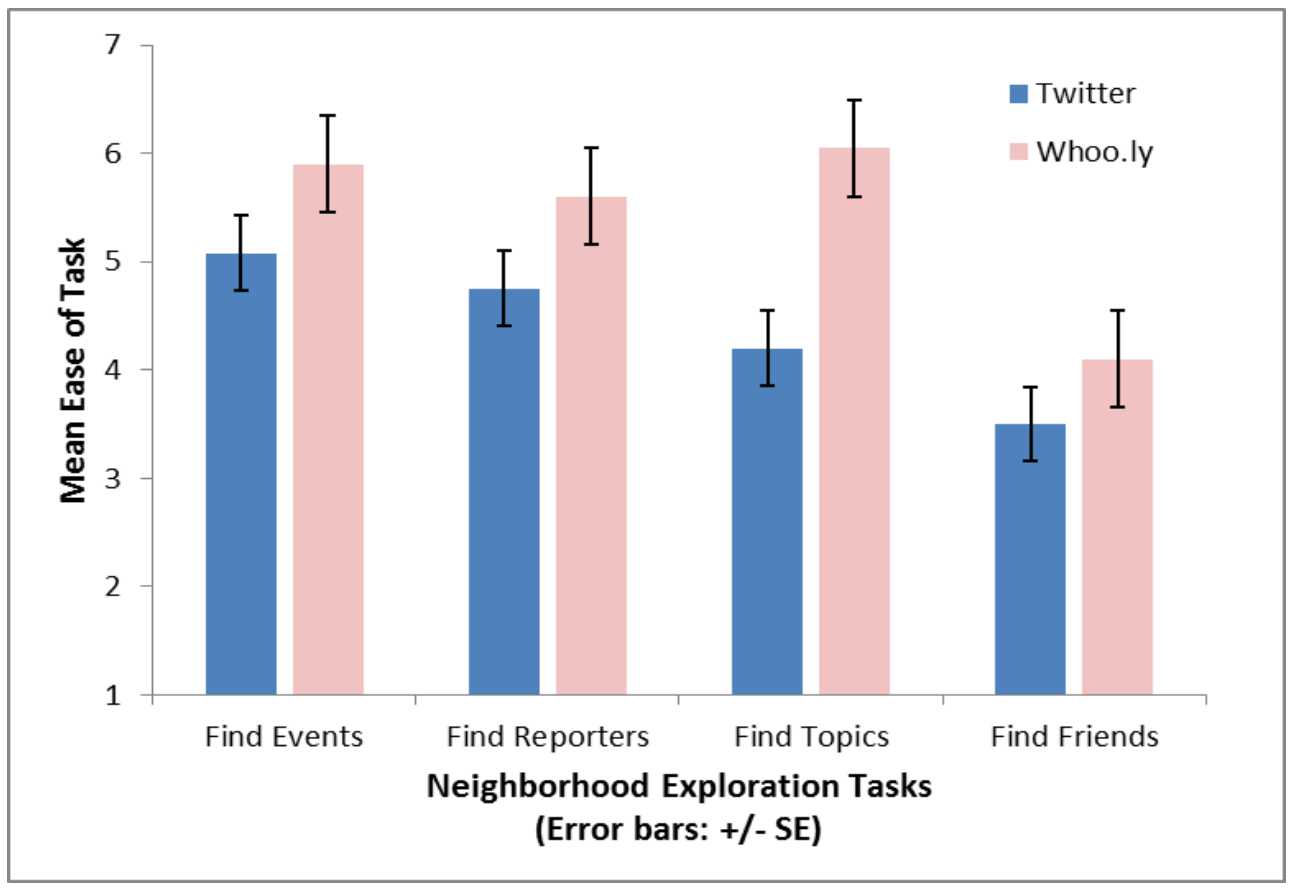

**Figure 3: Participants generally found it easier to complete neighborhood exploration tasks using Whoo.ly (where 1 = not at all, and 7 = extremely so.)**

Participants also completed overall ratings of Whoo.ly and Twitter, and again using an omnibus repeated measures ANOVA (technology X type of rating) we found an effect of technology ($F$(1,11) = 3.09, $p$ = 0.06), such that participants reported it as more useful ($F$(1,11) = 2.24, $p$ = 0.08) and easier to use ($F$(1,11) = 2.72, $p$ = 9.07), that it provided a better overview ($F$(1,11) = 2.74, $p$ = 0.07, and that it increased the sense of connection to their neighborhood community (F(1, 11) = 3.5, $p$ = 0.04), as shown in Figure 4. Again, neighborhood communication self-efficacy had a marginally significant interaction such that people with lower levels self-efficacy were more impacted by Whoo.ly in their ratings of sense of connection *(F(1, 11)* = 2.81, $p$ = 0.09).

To assess our event detector, we compared user ratings of 503 Twitter posts in our participants' neighborhoods to the event detectors. Users indicated that 170 of the total Twitter messages were event-related. Among these, the detector also identified 78% of messages as event-related, relative to 17% false positives. A logistic regression shows a strong, significant correspondence (*beta* = 0.53, $p$ < 0.001). The event detector also produced a score for the importance to prioritize events in the user interface, and this score was much higher for Twitter messages the participants identified as events ($t$ = 16.92, $p$ < 0.001). The participants' ratings of the importance of an event was significantly correlated with the event detectors ($r$ = 0.31, $p$ < 0.001).

In order to compare the relative value of the types of summarization provided by Whoo.ly, we asked participants to rank the five main sections by order of preference, where 1 = most preferred and 5 = least preferred. We found that participants rated recent events most highly ($M$ = 1.6), followed by the Tweet stream ($M$ = 2.8), the top topics (M = 3.2), active people ($M$ = 3.5), and popular places ($M$ = 3.5).

After participants completed both sets of tasks, we asked them to choose which application they would prefer to use to find out what is happening in their neighborhood. Eight participants out of 13 preferred Whoo.ly. However, when asked to compare it to their favorite neighborhood blog, eight out of 13 said they would prefer their neighborhood blog. On average, participants indicated they were somewhat likely to actually use Whoo.ly if it were made publicly available ($M$ = 4.4, $SD$ = 1.62 where 1 = not at all, and 7 = extremely so).

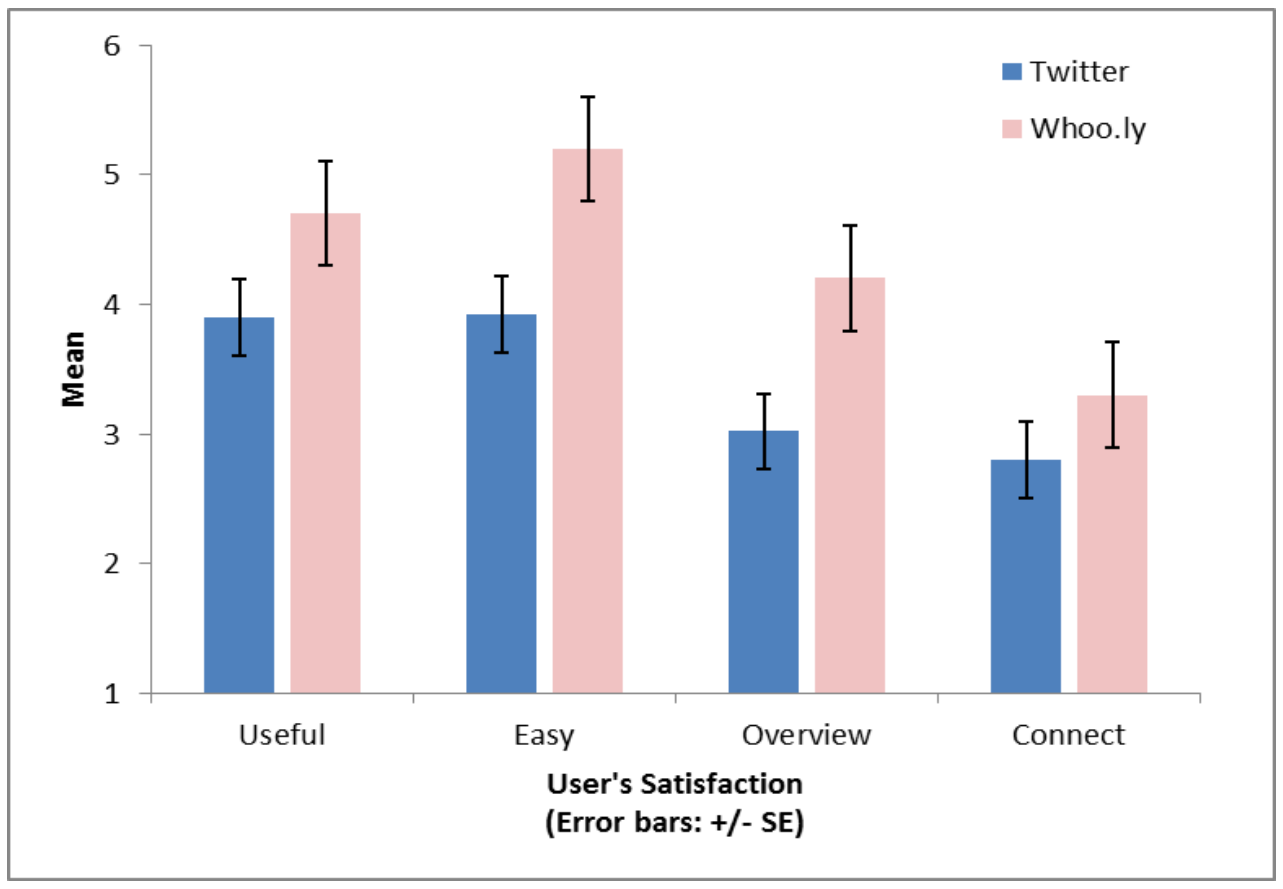

**Figure 4: Whoo.ly was found to be more useful, easy to use, with a better overview of the users' neighborhoods, and a sense of connection to their neighborhood communities.**

In order to shed light on some of our more quantitative findings, each participant was asked to provide feedback in writing about what they liked and disliked about Whoo.ly and how they would suggest improving it. Then, participants were asked to briefly discuss their experiences. When asked what they most liked about Whoo.ly, participants indicated the summarization and community features. Some of participants' answers were,

*"Really liked it overall, definitely a lot easier to find stuff"*

*"Whoo.ly was set up specifically with the community in mind. It makes community news/events/issues/people etc. easily accessible".*

When asked what they disliked, a few participants noted that a lot of the content felt like spam, such as the Craigslist postings, Foursquare check-ins, or overly personal posts, which interfered with participants' ability to access meaningful content. One participant said, *"Results. Mostly the furniture on craigslist. Need to filter out those, and be able to differentiate between the spammy 'top users' and the real top users."*

During the discussions, there were also several requests for further, personalized filters, to focus on the kind of content they cared about.

When asked why they preferred Whoo.ly over Twitter, again participants noted the filtering, summarization, and community features. One participant's answer was,

*"Twitter isn't set up for a community. Whool.y functions amazingly for this."*

Consistent with our more quantitative findings, we found that participants who preferred Twitter over Whoo.ly did so because they were already well-connected to their neighborhoods and already using Twitter to follow neighborhood reporters. For example, a participant's reply was, *"If I didn't know my neighborhood as well I would use both and compare data. Since I am very embedded in my community Whoo.ly is just another aggregator."*

When asked why they would prefer their local blog over Whoo.ly, participants noted blogs had more extensive features such as calendars and they benefitted from social curation. When asked why they would prefer Whoo.ly, participants mentioned its ease consumption and community feel. Some of the participants' reasons were,

*"like that it's short messages…easier than if browsing full blog with full messages; easier to figure what's going on."*

*"Whoo.ly offers not only news/events, but also connects you with people. Like combining Twitter and a newspaper, I like it".*

## DISCUSSION

As shown above, the overall reaction to the information provided on Whoo.ly was quite positive. The participants to our study found Whoo.ly easier to use than Twitter and the majority said they would prefer it as a tool for exploring their neighborhoods.

As a prototype system, Whoo.ly has advanced the state of the art for information seeking in hyperlocal communities, but many opportunities for improvement remain. As people cross the line from consuming hyperlocal information to engaging with their local community, they seek to know as much about the people as about the news. Thus, from a hyperlocal community perspective, it is also important to recommend potential similar friends such as "people like me in my neighborhood" as a feature to improve neighborhood connections. Besides, exploring the sentiments behind people's response/reactions to neighborhood issues can be useful [16]. Furthermore, it is interesting to note the unique characteristic of consuming social media when embedded in a geographical location—people could easily walk out their front doors and down the street to experience, for example, the local event they had just read about online.

It is worth noting that we deliberately placed the questionnaire and the focus group prior to the information-seeking user tasks to frame the tasks specifically on neighborhood seeking behaviors. Our intention was to give users the opportunity to have access to each other's neighborhood seeking experiences in evaluating the technology's effectiveness. We recognized a discussion could have systematically and artificially affected preferences towards or against Whoo.ly across all participants. However, there is no indication that this is the case. To further assess potential discussion confound, we tested for group size (2 vs. 3) on preference for Whoo.ly vs. Twitter, and found no effect. Moreover, we also found there were no session and level of Twitter usage effects.

## CONCLUSION

Whoo.ly is a web service that facilitates information seeking in hyperlocal communities by finding and summarizing neighborhood Twitter messages. In this paper, we presented several computational approaches used in Whoo.ly to discover hyperlocal content from noisy and overwhelming Twitter posts. In particular, we developed a novel event detector to discover trending events from recent posts. In addition, activity-based ranking algorithms and information extractors provided additional insights into the most active people and popular places in a local community. We performed a user study to evaluate Whoo.ly, and we found that (1) our event detector accurately identified events and (2) the local residents who participated in our study found Whoo.ly to be an easier tool for finding hyperlocal information than Twitter.

Social media such as Twitter has altered society's information and communication fabric and will continue to be increasingly integrated in our daily lives. We believe this paper presents a promising approach to leveraging Twitter messages to better support hyperlocal community awareness and engagement.